\documentclass{ws-procs10x7}

\begin{document}

\title{Coherent multiple scattering and dihadron correlations in heavy ion
 collisions}

\author{Jianwei Qiu}

\address{Department of Physics and Astronomy, Iowa State University,
Ames, IA 50011, USA \\E-mail: jwq@iastate.edu}

\author{Ivan Vitev}

\address{Los Alamos National Laboratory, Mail Stop H846,
Los Alamos, NM 87545, USA\\
E-mail: ivitev@lanl.gov}  

\twocolumn[\maketitle\abstract{
We present a systematic calculation of the {\it coherent} multiple
parton scattering with several nucleons in lepton-nucleus and  
hadron-nucleus collisions. We show that in $\ell+A$ reactions
coherence  leads to nuclear shadowing in the structure functions 
and a modification of the QCD sum rules. In $p+A$ reactions we 
evaluate the nuclear suppression of single and double inclusive 
hadron production at moderate transverse momenta. We demonstrate 
that both spectra and dihadron correlations in $p+A$ collisions at 
RHIC are sensitive measures of such dynamical nuclear attenuation 
effects.
}]


\section{Introduction}

Copious experimental data\cite{QM2004} from central $Au+Au$ 
reactions at the Relativistic Heavy Ion Collider (RHIC)
has generated tremendous excitement by pointing to the
possible creation of a deconfined state of QCD with energy 
density as high as 100 times normal nuclear matter 
density\cite{Vitev:2002pf}. In order to better understand
the jet quenching mechanism in $A+A$ collisions
we first need to address the multiple scattering between a partonic 
probe and the medium in simpler strongly interacting 
systems, like $\ell+A$ and $p+A$. 

In this talk we present a systematic calculation
of the {\it coherent} multiple parton scattering with several 
nucleons in lepton-nucleus and hadron-nucleus collisions. Recent 
theoretical developments\cite{Qiu:2003vd,Qiu:2004qk,Qiu:2004da}
have been able to provide a consistent picture of the dynamical
nuclear shadowing of sea quarks, valence quarks and gluons. 
Coherence leads to small $x \leq 0.1$ suppression\cite{Adams:1995is} 
of the structure functions $F_1^A$, $F_2^A$  and  $F_3^A$ in $\ell+A$ 
reactions\cite{Qiu:2003vd,Qiu:2004qk} and
suppression\cite{Arsene:2004ux} of single 
and double inclusive hadron production cross sections  at small 
and moderate transverse momenta in $p+A$ reactions\cite{Qiu:2004da}. 
Coherent multiple scattering and the well-studied elastic multiple
scattering\cite{Gyulassy:2002yv} co-exist in nuclear 
collisions and their relative role 
depends on the probes (or observables) and the underlying dynamics.
The interplay between these two scattering channels is beyond the
scope of this talk.


\section{Coherence and Shadowing}


\begin{figure}
\begin{center}
\includegraphics[width=1.2in]{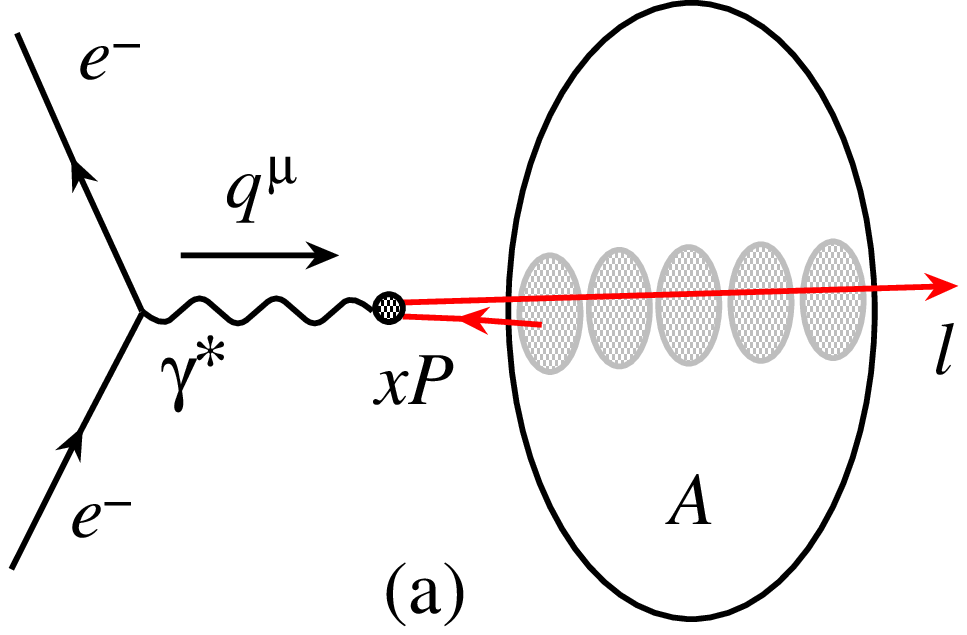}
\includegraphics[width=1.3in]{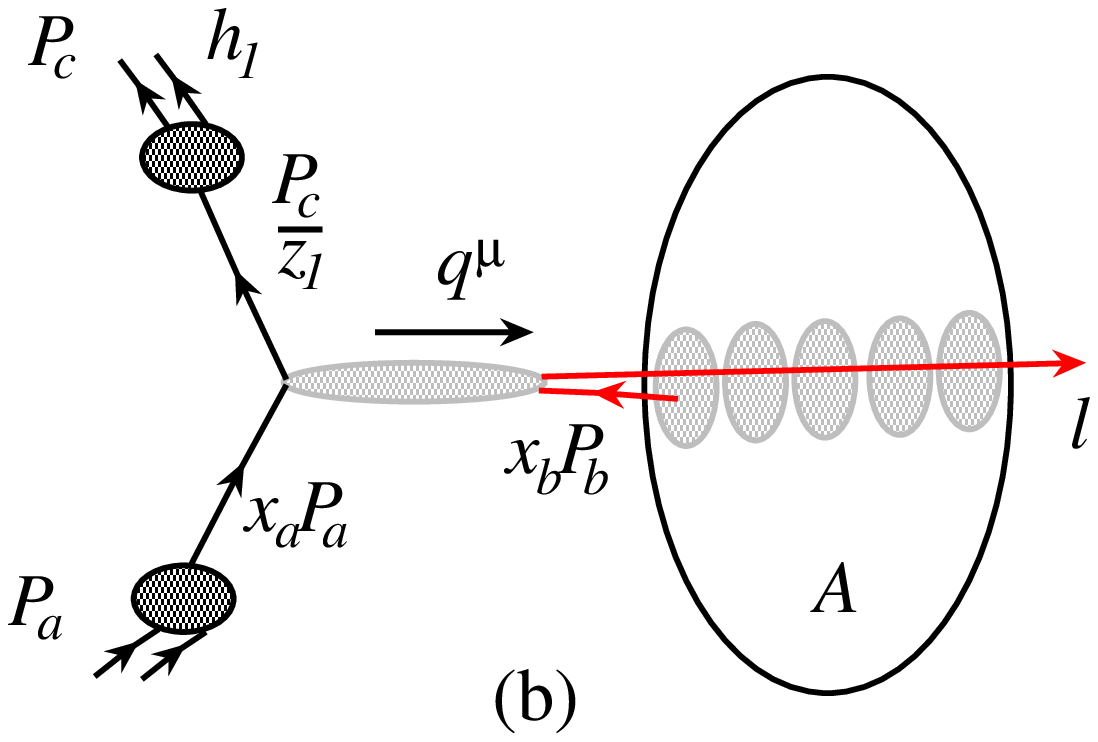}
\caption{From\protect\cite{Qiu:2004da}. 
Coherent multiple scattering of 
the struck parton in deeply inelastic scattering~(a) 
and in hadron-nucleus  collisions~(b). }
\label{fig1}
\end{center} 
\end{figure}



\begin{figure}[t!]
\begin{center}
\includegraphics[width=2.5in,height=3.8in]{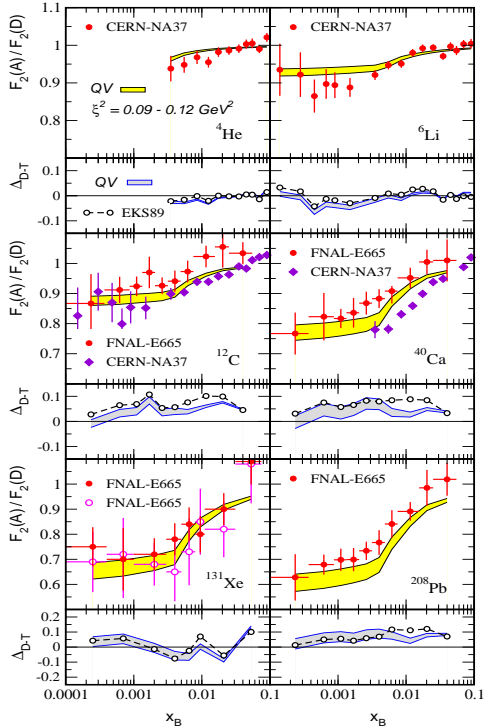} 
\caption{From\protect\cite{Qiu:2003vd}. 
All-twist resummed $F_2(A)/F_2(D)$ calculation 
versus DIS data\protect\cite{Adams:1995is}  
on nuclei. Data-Theory $\Delta_{D-T}$ is also shown. }
\label{fig2}
\end{center} 
\end{figure}


Hard scattering in nuclear collisions requires one large momentum
transfer $Q \sim xP \gg \Lambda_{QCD}$ with parton momentum 
fraction $x$ and beam momentum $P$. A simple example, shown in 
Fig.~\ref{fig1}(a), is the lepton-nucleus deeply inelastic scattering
(DIS).  The effective longitudinal interaction length probed by the
virtual photon of momentum $q^\mu$ is characterized by $1/xP$. If the
momentum fraction of an active 
parton $x \ll x_c=1/2m_N r_0\sim 0.1$  
with nucleon mass $m_N$ and radius  $r_0$, it could cover several  
Lorentz contracted nucleons  of longitudinal size $\sim 2r_0 (m_N/P)$
in a large nucleus.  In the photon-nucleus frame, 
Fig.~\ref{fig1}(a), the scattered parton of momentum $\ell$ 
will interact {\it coherently} with the partons from different
nucleons at the same impact parameter\cite{Qiu:2003vd}. 
In terms of the collinear factorization approach\cite{Collins:gx}, 
we showed\cite{Qiu:2003vd} 
that the hard partonic interactions
for any number of coherent multiple scattering in DIS are infrared
safe. The on-shell condition for $\ell$  fixes the  incoming parton's
momentum fraction to the Bjorken variable $x = x_B=Q^2/(2P \cdot q)$
but additional final state scattering forces the
rescaling\cite{Qiu:2003vd}  
\begin{equation}
\frac{1}{A} F_1^A( x_B ) = F_1 \left( x_B 
\left[1+ \frac{\xi^2 (A^{1/3}-1)}{Q^2} \right] \right) \; ,
\end{equation}
with the scale of power corrections 
$\xi^2\approx \frac{3\pi\alpha_s(Q)}{8r_0^2}
\lim_{x\rightarrow 0} \frac{1}{2} x G(x,Q^2)$.    
For $\xi^2=0.09-0.12$~GeV$^2$ the calculated reduction 
in the DIS structure functions, known as nuclear shadowing, is 
consistent with the $x_B$-, $Q^2$- and $A$-dependence of 
the data\cite{Adams:1995is}  as demonstrated in Fig.~\ref{fig2}.

Neutrino-nucleus DIS is a particularly instructive example since 
it provides the possibility to study the interplay between the 
dynamical power corrections and the heavy quark mass and is
a unique handle on valence quark shadowing. For 
$\nu + s \rightarrow   \ell^- + c$, 
the rescaling  of the Bjorken-$x$ reads:
\begin{equation}
x_B \rightarrow x_B \left( 1+  \frac{M_c^2}{Q^2} +  
 \frac{\xi^2 (A^{1/3} -1)}{Q^2} \right) \;, 
\label{StatDynShift} 
\end{equation}       
where $M_c$ is the charm quark mass. From Eq.~(\ref{StatDynShift}) 
it is evident that the role of the coherent multiple final state
scattering is to generate a dynamical parton mass
$m^2_{\rm dyn} = \xi^2 (A^{1/3} -1) $ in the nuclear 
chromomagnetic field\cite{Vitev:2004kd}. The rescaling of the   
value of $x_B$ reflects the conversion of the parton's energy 
into mass and correspondingly reduces the partonic flux. This 
attenuation depends on the $x-$slope of the parton distributions 
and is different for sea quarks and valence quarks\cite{Qiu:2004qk}.


\section{Suppression in Hadron-Nucleus Collisions}


Unlike in DIS, all diagrams with either final-state and/or
initial-state multiple interactions in hadron-nucleus collisions,
shown in Fig.~\ref{fig1}(b), could in principle lead to medium size
enhanced power corrections.  
However, once we fix the momentum fractions $x_a$ and
$z_1$, the effective interaction region is determined by the momentum
exchange $q^{\mu}=(x_aP_a-P_c/z_1)^{\mu}$.  In the head-on frame of
$q-P_b$, the scattered parton of momentum $\ell$ interacts coherently
with partons from different nucleons at the same impact parameter. 
Interactions that have taken place between the partons from the 
nucleus and the incoming parton of momentum $x_aP_a$ and/or the 
outgoing parton of momentum $P_c/z_1$ at a different impact parameter
are much less coherent and actually dominated by the independent
elastic scattering\cite{Gyulassy:2002yv}. 

Similarly to the DIS case\cite{Qiu:2003vd}, we find\cite{Qiu:2004da}
that resumming the coherent scattering with multiple nucleons is
equivalent to a shift of the momentum fraction of the active parton
from the nucleus in Fig.~\ref{fig1}(b),
\begin{equation}
x_b\rightarrow x_b 
\left (1+ \frac{C_d \xi^2(A^{1/3}-1)}{(-t)} \right) \, ,
\label{shift-pa}
\end{equation}
with the hard scale $t=q^2=(x_aP_a-P_c/z_1)^2$ and the color factor
$C_d$ depending on the flavor of parton $d$. $C_{q(\bar{q})}=1$ and 
$C_g=C_A/C_F=9/4$ for quark (antiquark) and gluon, respectively, since 
the gluons couple twice as strongly to the medium via their 
average squared color charge. 
The shift in Eq.~(\ref{shift-pa}) leads to a net suppression of the
cross sections, and the $t$-dependence of this shift indicates that
the attenuation increases in the forward rapidity region.


\begin{figure}
\begin{center} 
\includegraphics[width=2.5in,height=2.4in]{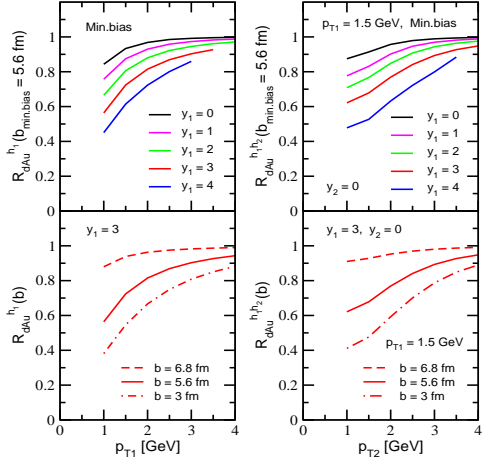}
\caption{From\protect\cite{Qiu:2004da}. Top: suppression of the single 
and  double inclusive 
hadron production rates in $d+Au$ reactions versus $p_T$ for 
different rapidities. 
Bottom: the calculated nuclear modification at different impact
parameters.  The trigger hadron $p_{T1}=1.5$~GeV,   
$y_1 = 3$ and the associated hadron  $y_2=0$. }
\label{fig2-mod-12}
\end{center} 
\end{figure}


Dynamical nuclear effects in multi-particle production can be 
studied through the ratio
\begin{equation}
R^{(n)}_{AB}  =  
\frac{\frac{d\sigma^{h_1 \cdots h_n}_{AB}}
           {dy_1 \cdots dy_n d^2p_{T_1} \cdots d^2p_{T_n}}} 
     {\langle N^{\rm coll}_{AB} \rangle\, 
      \frac{d\sigma^{h_1 \cdots h_n}_{NN}} 
           {dy_1 \cdots dy_n d^2p_{T_1} \cdots d^2p_{T_n}}} \, .
\label{multi}
\end{equation}
Centrality dependence is implicit in Eq.~(\ref{multi}) and the 
modified cross section per average collision 
$d\sigma_{dAu}^{h_1 \cdots h_n}/\langle N^{\rm coll}_{dAu} \rangle $ 
can be calculated\cite{Qiu:2004da} from the nucleon-nucleon 
cross section with the shift in Eq.~(\ref{shift-pa}). 
We consider $d+Au$ reactions at RHIC at $\sqrt{S}=200$~GeV in the
following discussions.   

The top panels of Fig.~\ref{fig2-mod-12} show the rapidity and
transverse momentum dependence of $R^{h_1}_{dAu}(b)$ 
and $R^{h_1h_2}_{dAu}(b)$ for minimum bias collisions. 
The amplification of the suppression effect at forward $y_1$ comes 
from  the steepening of the parton distribution 
functions at small $x_b$ and the decrease of the Mandelstam variable 
$(-t)$. At high $p_{T1},p_{T2} $ the attenuation is found to  
disappears in accord with the QCD factorization 
theorems\cite{Collins:gx}.  The bottom panels of Fig.~\ref{fig2-mod-12} 
show the growth of the  nuclear attenuation effect with centrality.

Dihadron~correlations 
\ $C_2(\Delta \varphi) = 
\frac{1}{N_{\rm trig}} \frac{dN^{h_1h_2}_{\rm dijet}}{d\Delta \varphi} $ 
associated with $2 \rightarrow 2$
partonic hard scattering processes, after subtracting the bulk
many-body collision background, can be approximated by near-side
and away-side Gaussians. The acoplanarity, $\Delta \varphi  \neq \pi$,
arises from high order QCD corrections and in the presence of nucler
matter - transverse momentum diffusion\cite{Gyulassy:2002yv}.  
If the strength of the away-side correlation function in elementary 
N+N collisions is normalized  to  unity, dynamical quark and gluon 
shadowing in $p+A$ reactions will be manifest in the  attenuation of  
the {\em area}  $A_{\rm Far} = R^{h_1h_2}_{pA}(b)$\cite{Qiu:2004da}. 


\begin{figure}
\begin{center} 
\includegraphics[width=2.5in,height=1.9in]{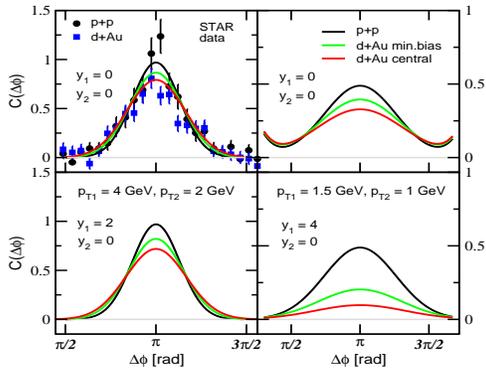}
\caption{From\protect\cite{Qiu:2004da}. Centrality dependence of 
$C_2(\Delta \varphi)$ at various rapidities and moderate (left) 
and small (right) transverse momenta.   
Central $d+Au$ and $p+p$ data  from STAR\protect\cite{Adams:2003im}.}  
\label{fig2-corr-fun}
\end{center} 
\end{figure}


The left panels of Fig.~\ref{fig2-corr-fun} 
show that for $p_{T_1}=4$~GeV, $p_{T_2}=2$~GeV 
the dominant effect in $C_2(\Delta \varphi)$ is a small increase 
of the broadening with centrality, compatible with the 
PHENIX\cite{Rak:2004gk} and STAR\cite{Adams:2003im} 
measurements. Even at forward rapidity, such as $y_1 = 2$, the 
effect of power corrections in this transverse momentum  range is  
not very significant. At small $p_{T_1}=1.5$~GeV,  $p_{T_2}=1$~GeV, 
shown in the right hand side of Fig.~\ref{fig2-corr-fun}, the apparent   
width of the away-side  $C_2(\Delta \varphi)$ is larger. 
In going from  midrapidity,  $y_1 = y_2 = 0$, to forward  rapidity, 
$y_1 = 4, y_2 = 0$,  we find a significant reduction by a factor 
of 3 - 4  in the strength of dihadron correlations. 
Preliminary STAR results\cite{Ogawa:2004sy} 
are consistent with our predictions.


\section{Conclusions}


In conclusion, we presented a systematic 
approach\cite{Qiu:2003vd,Qiu:2004qk,Qiu:2004da} 
to the calculation
of coherent QCD multiple scattering  and resummed the
nuclear enhanced power corrections to the structure functions
$F_1^A$,  $F_2^A$ and $F_3^A$ and to the single and double inclusive 
hadron production cross sections in $p+A$ reactions. At low 
$Q^2$ or $-t \propto p_T^2 $ 
we find a sizable suppression, which grows with centrality and 
via $-t$ and $x_b$ with rapidity. At high $Q^2$ ot $-t$ the nuclear 
modification disappears in accord with the QCD factorization 
theorems\cite{Collins:gx}. 
We demonstrated that both particle spectra and dihadron correlations 
are sensitive measures of such dynamical attenuation  of the parton 
interaction rates in $p+A$ reactions.

Our approach, with its intuitive and transparent results, 
can be easily applied to study the nuclear modification of other 
physical observables in $p+A$ reactions. The systematic incorporation 
of coherent power corrections provides a novel tool to address the 
most interesting transition region  between ``hard'' and ``soft'' 
physics in hadron-nucleus collisions.


This work is supported in part by the US Department of Energy  
under Grant No. DE-FG02-87ER40371 and by the J.R. Oppenheimer 
fellowship of the Los Alamos National Laboratory.



\end{document}